\documentclass[fleqn,10pt]{wlscirep}
\usepackage[utf8]{inputenc}
\usepackage[T1]{fontenc}
\usepackage{amsmath}

\DeclareMathOperator*{\argmax}{argmax}

\title{Multiscalarity in Socio-Spatial Segregation: An Information-Theoretic Framework}

\author[1,2,*]{Mateo Neira}
\author[1]{Valentina Marin}
\author[1]{Elsa Arcaute}
\affil[1]{Centre for Advanced Spatial Analysis. University College of London, London, W1T 4TJ, UK}
\affil[2]{Alan Turing Institute, British Library, London, NW1 2DB, UK}

\affil[*]{mateo.neira.16@ucl.ac.uk}

\begin{abstract}
We present a novel analytical framework to examine socio-spatial segregation across multiple spatial scales, explicitly leveraging information theory and percolation theory. This framework emphasizes the interplay between regional connectivity and population distribution, which are critical for understanding how spatial inequalities arise and persist in urban regions. Employing the Generalised Jensen-Shannon Divergence (GJSD), this method identifies regions characterized by significant segregation and low connectivity, providing actionable insights for targeted urban interventions. Using Ecuador as a case study, we demonstrate how segregation patterns manifest differently at city, regional, and national scales, underscoring the critical role of multiscalarity in understanding urban inequalities and guiding scale-sensitive policies. This approach not only advances the methodology for studying socio-spatial segregation but also contributes to the broader field by highlighting the importance of considering multiscalar perspectives and connectivity in urban systems.
\end{abstract}
\begin{document}

\flushbottom
\maketitle

\thispagestyle{empty}

\section*{Introduction}

    Socio-spatial segregation is a phenomenon with wide-ranging effects, ranging from physical to political and economic isolation. For example, limited access to healthcare in segregated neighbourhoods can result in systematic health disparities \cite{larrabee2022racialized}, while political isolation may diminish community representation \cite{newman2016diversity}. Despite abundant socio-economic data and extensive research on socio-spatial segregation \cite{massey1988dimensions, reardon20022, lee2008beyond, wong2011measuring}, significant challenges remain in capturing how segregation manifests differently across spatial scales. While extensive data is available, it often addresses singular scales of analysis; bridging local and global perspectives requires methodologies that capture spatial dynamics across multiple scales. This ongoing challenge drives the continuous evolution and improvement of methodological approaches, as observed in the literature \cite{reardon2009race,clark2015multiscalar, fowler2016contributions, petrovic2018multiscale, chodrow2017structure, olteanu2019segregation}. 

    Multiscalarity in urban systems refers to the analysis and understanding of socio-spatial phenomena across different spatial scales. It encompasses levels ranging from local neighbourhoods to entire metropolitan regions or broader geographic extents, emphasizing how processes manifest differently depending on the level of observation \cite{rozenblat2018conclusion}. This concept is crucial because patterns of socio-spatial segregation can appear homogeneous at a city-wide scale yet reveal significant disparities at the neighbourhood level or between urban subregions \cite{chodrow2017structure}. Understanding these varying dynamics offers policymakers a more comprehensive picture of urban systems \cite{marin2024scalar}. For example, city-wide economic policies aimed at addressing disparities may not align with regional transportation plans, leading to mismatches that fail to address local socio-economic challenges and highlight the need for scale-aware interventions. 

    Multiscalarity is fundamental for effective policy and decision-making because urban phenomena such as segregation, connectivity, and accessibility do not operate uniformly across scales. Policies crafted with a multiscalar perspective can better address the complex nature of urban challenges by recognizing that interventions may have varying impacts at different scales \cite{rozenblat2009european}. For instance, strategies to improve regional transportation connectivity can influence local socio-economic conditions, potentially reshaping segregation patterns \cite{barbosa2021uncovering}. Multiscalar analysis also helps identify where targeted interventions are needed most, ensuring that resources are allocated efficiently and where they can have the most significant impact. 

    The challenge of measuring socio-spatial segregation across scales lies in the fact that traditional approaches rely on predefined spatial units such as census tracts or neighbourhoods, potentially masking segregation patterns that emerge at different scales \cite{yao2019spatial}. Studies using census tracts may overlook broader regional exclusion patterns, as seen in research on metropolitan transportation zones where a narrow focus missed wider socio-spatial disparities \cite{chodrow2017structure}. While useful, these approaches may miss critical patterns of segregation that only emerge at certain scales. A city may show low overall levels of segregation when analysed as a whole, but detailed neighbourhood-level examinations might reveal significant disparities \cite{johnston2016macro}. Conversely, neighbourhoods that seem integrated may be part of broader city-wide segregation patterns. This highlights the necessity of a multiscalar approach that captures the full extent of socio-spatial segregation, enabling the identification of the scales where segregation is most pronounced and where interventions are most effective.

    To address these challenges, we present a new analytical framework to examine socio-spatial segregation across multiple spatial scales. This framework integrates information-theoretic measures with network-based spatial analysis, enabling both topological and spatial insights. The framework captures both the spatial distribution of populations and the topology of regional connectivity. This approach identifies the scales at which high segregation and low connectivity coincide, offering nuanced insights into socio-spatial dynamics. We apply our framework to Ecuador, a context with pronounced regional disparities, to demonstrate its capacity to identify segregation patterns and guide effective urban interventions. Our findings underscore the framework's potential to guide targeted urban interventions and support scale-aware policymaking.

\subsection*{Related works}

    Socio-spatial segregation refers to patterns in the spatial distribution of population groups that significantly diverge from random distributions \cite{winship1977revaluation}. These patterns span occupational, residential, and social dimensions and arise from social and spatial constraints. While individual choices influence their distributions, the aggregate effects often produce unexpected outcomes \cite{Bouchaud2013}.

    To measure social-spatial segregation, traditional segregation studies assess whether groups live together or apart within predefined spatial units. Aggregate-level indices, such as the Index of Dissimilarity \cite{duncan1955methodological} and Gini’s Coefficient \cite{gini1912variabilita}, quantify unevenness but overlook spatial proximity, treating segregation as "aspatial." Spatial indices addressed this by incorporating clustering and proximity, grouped into geostatistical methods \cite{wong1999geostatistics}, adaptations of aspatial measures \cite{reardon2004measures}, and smoothing kernels \cite{o2007surface}.

    In sociology, segregation can also be studied at the individual level by capturing interactions among individuals as a network and using descriptive statistics to look at the differences between connectivity patterns within groups and between groups. These network science approaches model connections between individuals, introducing measures like homophily \cite{lazarsfeld1954friendship} and assortative mixing \cite{newman2003mixing}. Recent research on quantifying segregation has extended these methods to capture segregation patterns in different settings, for example work-place or activity space segregation \cite{wong2011measuring}, as well as to tackle some of the challenges inherent to the spatial complexity of the phenomenon: this relates to scale interpretability, the modifiable area unit problem, location equivalence, and population density invariance \cite{rodriguez2016overview}.

    Building on both spatial and network methods, recent research has introduced random-walk-based metrics \cite{sousa2022quantifying, neira2024urban} that offer a novel approach to understanding segregation through dynamics within networks. In these works, segregation is defined in terms of probabilities of encounters. For example, Sousa and Nicosia introduced a measure that focuses on the Class Coverage Time (CCT), quantifying how long a random walk takes to encounter different population groups and providing a probabilistic measure of segregation. Neira et al. introduced a multilayered network approach, incorporating random walks and Lévy flights to model socio-spatial segregation within transport systems. Their method accounts for connectivity patterns, temporal constraints, and diverse modes of transit, quantifying interaction probabilities across layers of the network. By applying this framework to real-world cities, they demonstrate its utility in evaluating transport infrastructure's impact on segregation. Together, these methods provide robust, flexible measures of segregation that can handle complex urban networks and reveal the role of transport in shaping social interactions.

    Despite these advancements, many approaches examine segregation at a single spatial scale, which can obscure patterns that only emerge across multiple scales and fail to capture localized complexities. Multiscalar methods address this limitation by analysing how segregation patterns vary across different scales, defined by geographical distances \cite{reardon2009race}, neighbouring counts \cite{clark2015multiscalar}, or aggregate units like census tracts \cite{fowler2016contributions, chodrow2017structure}. 

    Reardon and colleagues’ analysis of racial residential segregation in the United States during the 1990s provided an essential foundation for multiscalar segregation research \cite{reardon2009race}. By decomposing segregation metrics across geographic scales, their work revealed that patterns of racial division could vary significantly depending on the spatial lens applied. This insight emphasized that neighbourhood-level diversity does not necessarily imply integration at broader scales, such as entire metropolitan areas. Their findings underscored the importance of examining segregation not merely as a static phenomenon but as a multilevel dynamic shaped by interactions between local and regional processes.

    In a similar vein, Clark et al.’s study of neighbourhood composition in Los Angeles from 2000 to 2010 used an innovative, location-based approach to demonstrate how segregation and diversity manifest differently across urban scales \cite{clark2015multiscalar}. By employing spatial association metrics and multilevel regression analyses, they demonstrated that demographic shifts at the block level often contrast with larger-scale trends. Their work provided a nuanced perspective on how diverse urban areas can remain segregated when viewed from a more expansive spatial frame, reinforcing the importance of capturing these interactions across multiple levels.

    Fowler and colleagues contributed further by examining how individual places within metropolitan regions influence overall changes in segregation patterns \cite{fowler2016contributions}. Their decomposition of diversity shifts across time and space highlighted that suburban growth, depending on its spatial configuration, can either reinforce or mitigate metropolitan segregation. This approach illuminated the complex interplay between expanding urban peripheries and core cities, offering new perspectives on the suburbanization of racial and socio-economic divides.

    Beyond traditional statistical measures, Chodrow introduced an entropy-based framework grounded in information theory to quantify spatial segregation \cite{chodrow2017structure}. This approach captured the degree of spatial clustering and heterogeneity across neighbourhoods without relying on pre-defined administrative boundaries. By using Bregman divergence to assess the “information content” of a city’s socio-spatial structure, Chodrow addressed the long-standing issue known as the Modifiable Areal Unit Problem (MAUP), wherein results are heavily influenced by the chosen geographic unit. This methodological innovation provided a more robust means of characterizing segregation, independent of arbitrary spatial divisions.

    Hennerdal extended these advances by proposing a clustering algorithm that adapts to population density and spatial distribution, further mitigating the effects of fixed spatial units \cite{hennerdal2017multiscalar}. His adaptive window approach adjusted the scale of analysis dynamically, reflecting local population characteristics and thereby avoiding the biases inherent in static geographic boundaries. This method marked an important step toward more context-sensitive measures of segregation, applicable across diverse urban environments.

    Despite these advances, key challenges remain. Determining the appropriate geographic scale for measuring segregation remains a central question, as different spatial levels can yield contradictory insights. While adaptive methods, such as those developed by Hennerdal, offer potential solutions, further research is needed to refine these approaches for broader application. Additionally, the temporal dimension of segregation—how socio-spatial divisions evolve over time—requires more comprehensive longitudinal analyses. Although Fowler’s work highlights temporal shifts, there is still a need for methods that capture continuous changes in segregation dynamics.

    Another critical issue is the integration of networked spaces into segregation measures. While traditional studies have focused on residential clustering, the rise of mobility data has prompted a shift toward understanding how transport networks and daily commutes influence segregation. Petrovic’s work \cite{petrovic2018multiscale} provides an important foundation, but future studies could deepen this approach by examining how transport accessibility shapes social interactions and reinforces or disrupts segregation patterns.

    Understanding socio-spatial segregation across scales and contexts requires integrating multiple methodological frameworks—from classical indices to network-based and random-walk approaches. However, the challenges identified, such as geographic scale, spatial boundaries, and temporal dynamics, highlight the need for more flexible and comprehensive approaches.

    The methodology proposed in this study addresses these challenges by introducing an analytical framework that leverages percolation theory to construct a hierarchical representation of urban systems. By employing the Generalised Jensen-Shannon Divergence (GJSD), the framework captures the information loss across scales, thereby quantifying changes in segregation patterns at multiple spatial levels. This approach addresses the limitations of static spatial units and arbitrary boundaries by dynamically modelling socio-spatial interactions through the integration of socio-economic data and network connectivity. It specifically responds to key limitations in existing multiscalar approaches by refining the geographic scale question, incorporating dynamics through percolation-based hierarchies, and addressing the challenges of integrating networked spaces into segregation measures. By doing so, it provides a more comprehensive way to capture how urban connectivity influences socio-spatial divisions across multiple scales.
    
    This framework not only reveals regions characterized by high segregation and low connectivity but also provides actionable insights into how urban infrastructure contributes to or mitigates spatial inequalities. By explicitly accounting for multiscalar dynamics the proposed method underscores the importance of considering hierarchical urban structures in the study of socio-spatial segregation.

\section*{Methods}

    \begin{figure}[hbt!]
    \includegraphics[width=1\linewidth]{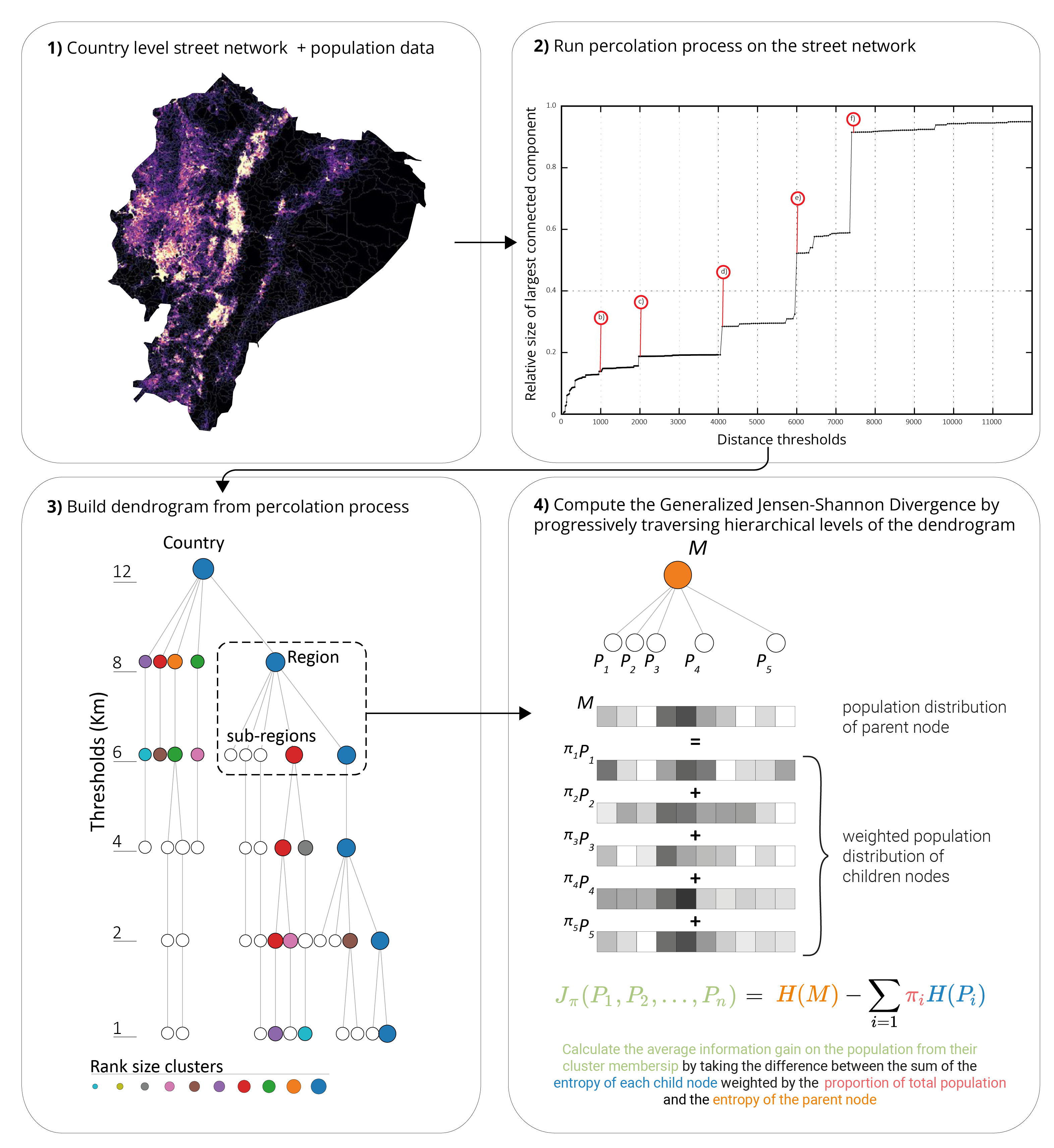}
    \caption[General overview of the proposed method to calculate multiscale segregation using Generalised Jensen-Shannon Divergence]{General overview of the proposed method to calculate multiscale segregation using Generalised Jensen-Shannon Divergence. A normalised GJSD measure is calculated (4) for every parent node in the dendrogram (3), defined by the percolation process (2) on the road network (1).}
    \label{fig:GSJD_method}

    \end{figure}
    
    Socio-spatial segregation can be defined as any pattern in the spatial distribution of population groups that deviates significantly from a random distribution. By taking an information theoretic view, measuring social spatial segregation can be formalised as calculating the degree of dissimilarity or deviation between the observed spatial distribution of different social groups and an expected random distribution. Through this information theoretic framework, we can conceptualise socio-spatial segregation by considering a set of spatial units $I$ and a set of population groups $B$ and a joint distribution $p_{I,B}$ in which $p_{I,B}(i,b)$ is the probability that an individual sampled uniformly at random from a region lives at location $i$ and is member of group $b$.
    
    Socio-spatial segregation can be quantified using the mutual information $\mathcal{I}(I; B)$, which measures the amount of information that knowing the location of an individual $I$ provides about their social group membership $B$, and vice-versa. Here, the total mutual information is the sum of the individual group contributions to the overall mutual information, mathematically defined as:
    
     \begin{equation}
          \mathcal{I}(I; B) = \sum_{i \in I} \sum_{b \in B} p_{I, B}(i, b) \log \left( \frac{p_{I, B}(i, b)}{p_I(i) p_B(b)} \right)
     \end{equation}
    where:  \\

    * $p_{I, B}(i, b)$ is the joint probability that an individual selected at random is located at spatial unit $i$ and belongs to group $b$.
    
    * $p_I(i) = \sum_{b \in B} p_{I, B}(i, b)$ is the marginal probability of being in location $i$.
    
    * $p_B(b) = \sum_{i \in I} p_{I, B}(i, b)$ is the marginal probability of belonging to group $b$.\\
    
    The mutual information measures the divergence of the joint distribution $ p_{I, B} $ from the product of the marginal distributions $p_I$ and $p_B$. When the mutual information is high, knowing the location $I$ of an individual provides significant information about their group membership $B$, indicating strong socio-spatial segregation.
    
    Using this framework as a starting point, we formulate the problem of measuring socio-spatial segregation across scales and identify the relevant scales at which both high segregation and low connectivity occur into the following question: How much information about an individual's group membership $b$ can we gain from knowing the area $i$ in which they reside and its connectivity to the wider region? 
    
    In order to answer this question we propose a framework that uses the hierarchical structure of urban systems and measures the average gain in information of population membership when regions are split into sub-regions. The framework can be summarised as follows and is represented in Figure \ref{fig:GSJD_method}:
    
    \begin{enumerate}
    \setlength\itemsep{-0.4em}
    \item Merge country level street network with population data.
    \item Run a percolation process on the street network and identify discontinuities at specific distance thresholds.
    \item Build a dendrogram from the percolation process where at the top we have a parent node that represents the entire country, which branches into smaller regions, systems of cities and conurbations, until it reaches individual cities as leaf nodes.
    \item Measure the average loss of diversity in population groups when regions are split into sub-regions by calculating the generalised Jensen-Shannon Divergence as specified in Eq.~(\ref{eq:GJSD}) (see next subsections) for each parent node and its child nodes within the dendrogram.
    \end{enumerate}

    The following subsections will provide a detailed description of each of the steps listed above: constructing the hierarchical structure of an urban system, calculating the distribution of population groups in the hierarchical tree, and measuring the average loss of diversity. Each subsection will provide clarity on the process and its relevance in measuring socio-spatial segregation.

    \subsection*{Multiscalarity in street networks}

        Given a graph $G = (N,L,w)$ of the road network, where nodes $N$ represent intersections and the weight $w_{i,j}$ for each link $L$ between nodes $i$ and $j$ is the length of the street that connects them, a network percolation process is defined such that for a given threshold $\delta$ we extract sub-graphs with connected links $w_{i,j}\leq \delta$. 
        
        We define a function $\varphi_{\delta}(i)$ for each node $i$ in the set $N$ and a given threshold $\delta$. This function assigns a label $k$ from the set $K_{\delta}$ to each node, where $k$ represents the connected sub-graph to which the node belongs. Specifically, if nodes $i$ and $j$ in $N$ are part of the same connected sub-graph when considering the threshold $\delta$, then $\varphi_{\delta}(i)$ and $\varphi_{\delta}(j)$ will be equal.
        
        The size of each group can be defined as $S_{\delta}(k) = | \{i\in N : \varphi_\delta(i) = k\} |$, which counts the number of nodes in each group $k$ for a given threshold $\delta$.
        
        Then, the size of the largest connected component for each threshold $\delta$ can be given as $|C_{\text{max}, \delta}| = \max\{S_{\delta}(k) : k \in K_{\delta}\}$, where $K_{\delta}$ is the set of all possible group memberships at threshold $\delta$.
        
        Analysing the largest sub-graph size across all thresholds can reveal scale transitions, which are indicated by discontinuities at specific thresholds. These thresholds represent the main levels of a hierarchical structure. Using these specific thresholds, a hierarchical tree is constructed where each node represents a $k$ group at the specified threshold $\delta$. At the top we have a parent node that represents the entire country, which branches into smaller regions, systems of cities and conurbations, until it reaches individual cities as leaf nodes.

    \subsection*{Distribution of population groups in hierarchical trees}

        In the percolation process described above we constructed a hierarchical tree that represents the multiscalar structure of our urban system. To measure segregation at each scale of this tree we need to integrate data about the underlying population. Such population data is usually gathered and aggregated at pre-defined spatial units, census tracks for example. Here, we define a procedure to aggregate population data from these pre-defined spatial units to each node of the hierarchical tree. 
        
        Consider a set of population groups $B$, with the unique property that each individual belongs to exactly one group. Let $n_{i, b}$ denote the count of individuals from group $b \in B$ residing in spatial unit $i \in  I$ defined by census data (such as census tracks or blocks).
        
        Our aim is to calculate a vector $\vec{P}_{k, \delta}$ for each node in the hierarchical tree $k \in K_{\delta}$ at all hierarchical levels $\delta$, where each element $p_{b, k, \delta}$ of this vector corresponds to the proportion of individuals from group $b$ associated with node $k$. It holds true that $\sum_{b \in B}p_{b, k, \delta} = 1$, where we can view $\vec{P}_{k, \delta}$ as encoding the probability of an individual in node $k$ to belong to different population groups $b$. 
        
        Let us now quantify the relative contribution of each child node $k$ to the population of its parent node in the hierarchical structure, by $\pi_{k} = \frac{n_k}{n}$, where $n_k$ is the total population of the child node $k$, and $n$ is the total population of the parent node.
        
        We start the process by assigning a group membership $k \in K_{\delta}$ to each spatial unit $i$ (census tracks). This assignment is based on the proportion of the population of the unit that belongs to each cluster from our hierarchical tree. This approach ensures a more accurate representation of the population distribution across clusters. Since the leaf nodes of our hierarchical trees are cities, and given that our census tracts are smaller subdivisions of these cities, we ensure a unique one-to-one mapping from our spatial units to the tree nodes. This method avoids double counting when advancing to the next level in the hierarchical tree.
        
        Following this, for each node $k \in K_{\delta}$ in the hierarchical tree, we calculate $n_b$ as the sum of all $n_{i, b}$ values for all $i$ that belong to $k$. Lastly, we calculate $n_{k}=\sum_{b}n_{b}$, the total population in $k$, and construct the vector $P_{k, \delta}$ by dividing each summed $n_{b}$ by $n_{k}$ . This effectively transforms our calculations into proportions of population distributions across different groups within each node in the tree.

    \subsection*{Multiscalar segregation on hierarchical trees}

        Once we have a hierarchical tree along with the population group distribution and population total calculated for each node in our tree we can address our original question: How much information about an individual's group membership $b$ can we gain from knowing the area $i$ in which they reside and its connectivity to the wider region? We can re-frame this with respect to the hierarchical tree by asking: Given a parent node in the tree, how much information about an individual's group membership $b$ can we gain by knowing to which child-node it belongs to? 
        
        Let us take the simplest case of a parent node only having two child nodes. We let $M$ be the population distribution of the parent node and $\pi_1$, $P_{1}$ and $\pi_2$, $P_{2}$ be the population percentage, and population distributions of each child node respectively. The average information content of each one of these probability distributions can be calculated through Shannon's entropy:
        
        \begin{equation}
        \label{eq:entropy}
        H(X) = -\sum_{x} p(x)\log p(x),
        \end{equation}
        hence for our parent node, the average information content would simply be $H(M)$. If we split the parent node into the two child nodes we can associate any individual to one of these nodes and, on average, reduce our uncertainty about its population group. The average information content of our population distributions is then given by the weighted average of the individual entropy's of each child node, $\pi_1 H(P_{1}) + \pi_2 H(P_{2})$. The average reduction in uncertainty can then be measured by the difference between the parent nodes' entropy and the average of the child nodes' entropy, which is equal to the Jensen-Shannon divergence \cite{lin1991divergence} between the two child nodes:
        
        \begin{equation}
        \label{eq:JSD_weighted}
            \text{JSD}_{\pi}(P_1 || P_2) = H\left(M\right) - \left[ \pi_1 H(P_1) + \pi_2 H(P_2) \right]
        \end{equation}

        We can generalise this for the case where a parent node splits into 2 or more child nodes with varying populations by using the Generalised Jensen Shannon Divergence (GJSD): 
        
        \begin{equation}
        \text{JSD}_{\pi}(P_1,P_2,...,P_n) = H(M) - \sum_i \pi_i H(P_i)
        \label{eq:GJSD}
        \end{equation}
        Note that when all child nodes have the same population groups $\text{JSD}_{\pi}$ is equal to zero. GJSD takes its maximum value when all of the $P_i$ are non-overlapping over their respective population groups. The upper bound of $\text{JSD}_{\pi}$ is determined by $n$, the number of distribution being compared. If we have $n$ completely distinct distributions, where each distribution assigns a probability of 1 to a unique population group not shared by any other distribution, then these distributions are maximally different. The upper bound of $\text{JSD}_{\pi}$ is given by:
        
        \begin{equation}
         JSD_{\text{max}} = -\sum_{k=1}^{n} \pi_k \log(\pi_k)
        \end{equation}
        
        We use this upper bound to normalise the $\text{JSD}_{\pi}$, for all nodes to allow comparisons for parent nodes with varying number of child nodes: $\text{JSD}_{\pi_{\text{norm}}}=\text{JSD}_\pi / \text{JSD}_{\text{max}}$. A formal derivation of this maximum is provided in the supplementary materials.
        
        With this, we have defined our multiscalar segregation measure through our hierarchical tree (see Figure \ref{fig:GSJD_method}). Each node in our tree has a $\text{JSD}_{\pi_{\text{norm}}}$ value associated with it (except for the leave nodes). By looking at each level of our hierarchical tree, we can identify the most segregated regions in a urban system at that particular scale. 
        
        Using these values we can traverse up through the tree starting at each leave node until we reach the root node. 
        Let $\text{JSD}_{\pi_{\text{norm}}}(i,\delta)$ denote the normalised GJSD of the node containing the leaf node $i$ at level $\delta$.  We can identify the most disconnected and segregated region through all scales for each leaf node (city) $i$ by finding the $\delta$ for which $\text{JSD}_{\pi_{\text{norm}}}(i,\delta)$ is maximised:
        
        \begin{equation}
          \Upsilon_{i} = \argmax\limits_{\delta} \left\lbrace
          \text{JSD}_{\pi_{\text{norm}}}(i,\delta) \right\rbrace
          \label{eq:segregation_scales}
        \end{equation}
        The variable $\Upsilon_{i}$ in our approach serves as a crucial metric for identifying individual cutoff points $\delta$ within the hierarchical tree. These cutoff points help in pinpointing clusters across various scales where socio-spatial segregation and disconnections are pronounced. This methodology provides a robust framework for quantitatively assessing and comparing segregation across different scales and regions within an urban system.

\section*{Results}

    We illustrate the method with a study of urban segregation for the country of Ecuador. Prior to any analysis relevant socio-economic groups must be defined for the population. To achieve this we use data available from the census to derive an index of life conditions at the household level and aggregate the index to urban blocks. The census data used is provided by the National Institute of Statistics and Census (INEC) of Ecuador. The Census data is available at the household level and is aggregated up to the block level. Following the calculation of an index of life conditions, the population is then divided into deciles, so that each population group contains the same amount of people, similar to the work by Dong et al. \cite{dong2020segregated}. Although traditional indices provided by the Ecuadorian government, such as the Unsatisfied Basic Needs (UBN) take into account monetary and non-monetary indicators, they divide the population only into two groups: above and below the poverty line. Because of this we adopt a combination of the index of life conditions (ILC) proposed by Sarmiento for the context of Colombia \cite{sarmiento1996indice}, and the Multidimensional poverty index for Latin America proposed by Santos and Villatoro \cite{santos2018multidimensional}. The former has the benefit of providing a continuous score across all dimension considered, while the latter can be used to compare across other countries in the region. The details of how the index is calculated can be found in the supplementary materials. 

    \subsection*{Hierarchical structure to Ecuador's transport network}

        \begin{figure}[hbt!]
            \centering
            \includegraphics[width=0.9\linewidth]{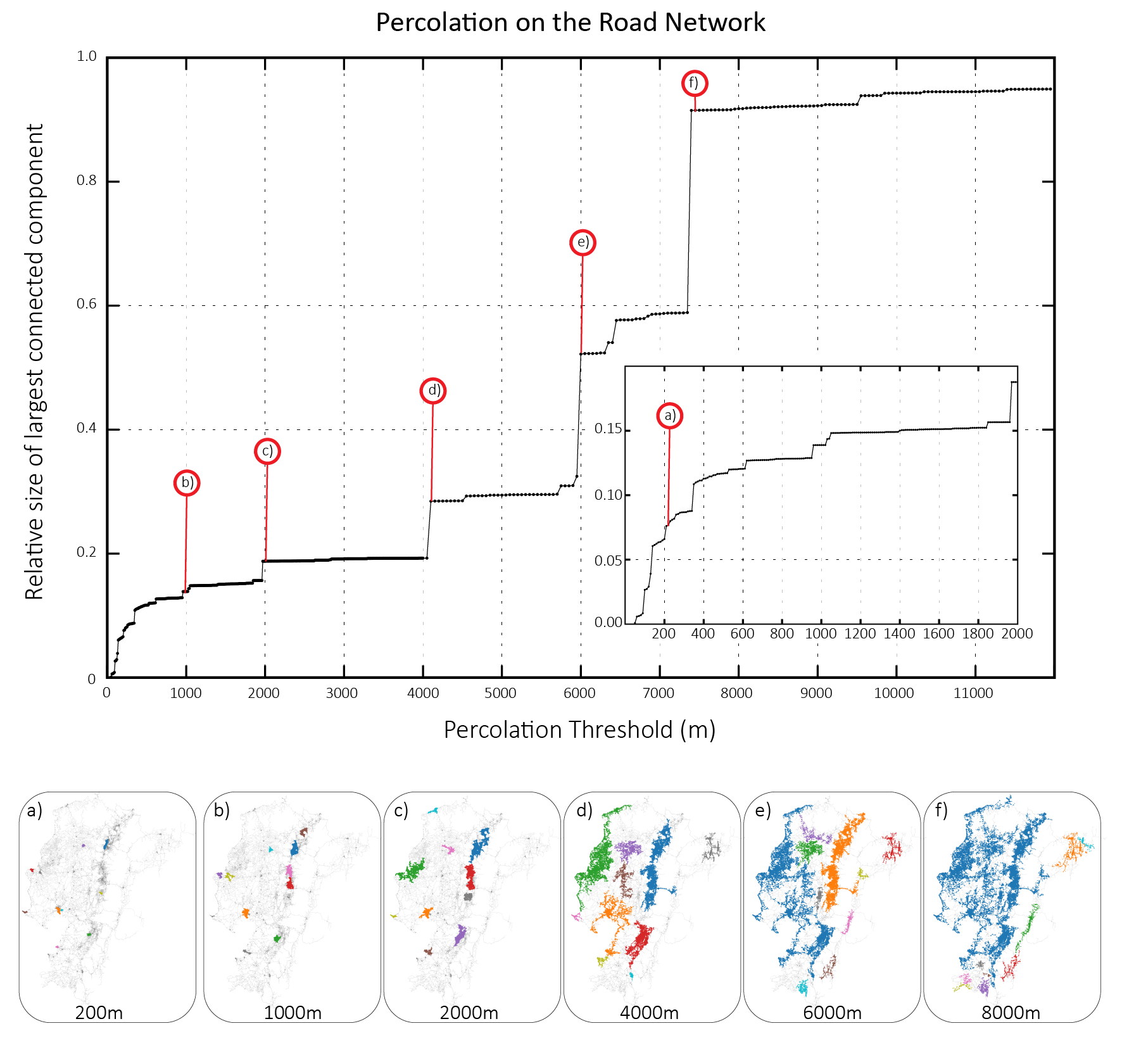}
            \caption[Evolution of the largest cluster size for the percolation on the street network of Ecuador.]{\label{fig:percolation} Evolution of the largest cluster size for the percolation on the street network of Ecuador. Maps show the 10 largest clusters at selected thresholds. Cities appear at around the $200m$ meter clusters. At the 6km threshold the three major regions (coast, highlands, and the Amazon) can be clearly distinguished.}
        \end{figure}

        \begin{figure}[hbt!] 
            \centering
            \includegraphics[width=0.9\linewidth]{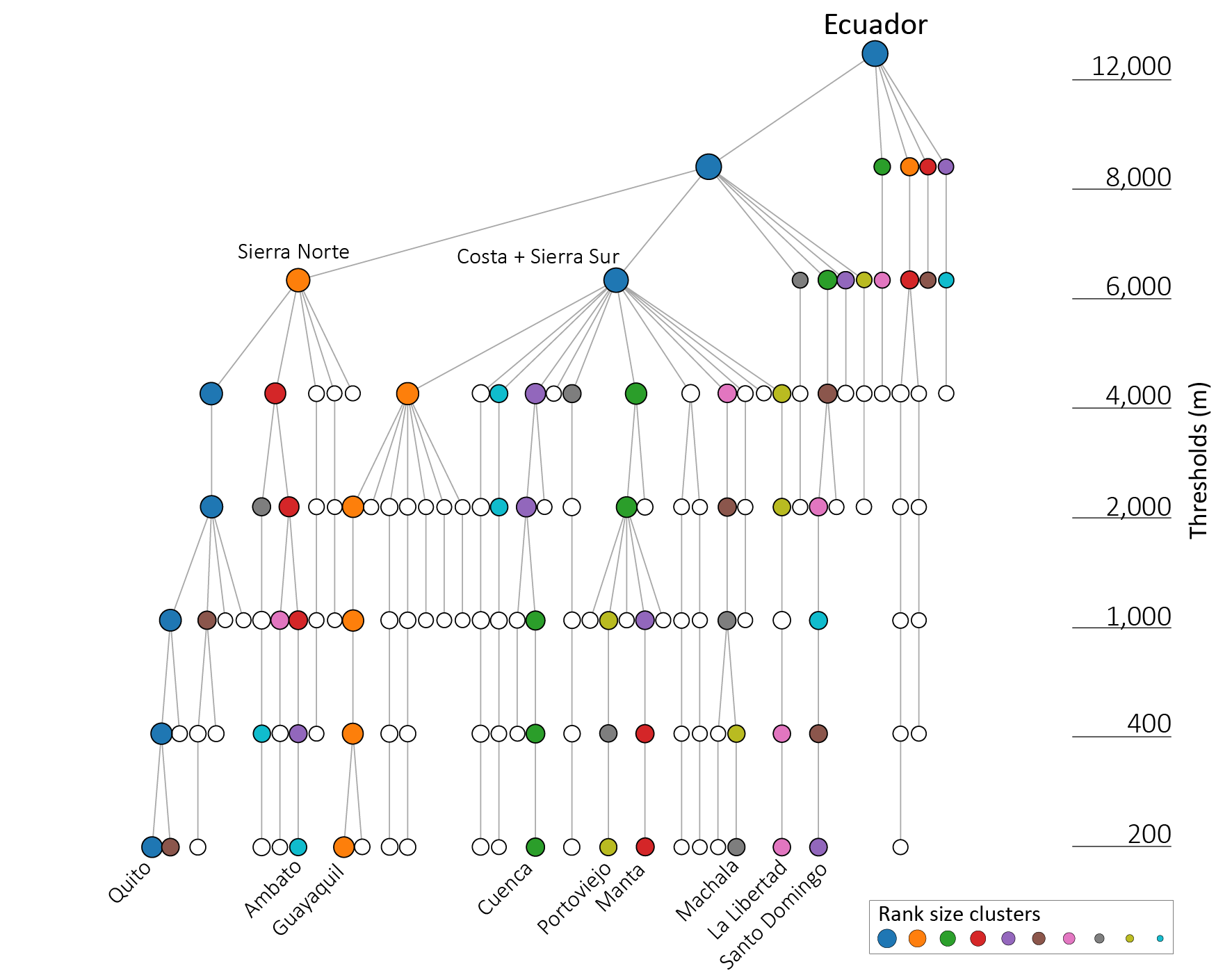}
            \caption[Hierarchical tree of the evolution of some of the largest clusters through the percolation process.]{\label{fig:hierarchical_tree} Hierarchical tree of the evolution of some of the largest clusters through the percolation process. The size is measured according to the number of nodes. It's important to note that the Southern Highlands before joining with the Northern Highlands}
        \end{figure}
    
        To study the hierarchical structure of the road network of Ecuador and derive relevant spatial scales for the analysis, we build a road network graph for the whole country using OpenStreetMap data pertaining to all drivable roads. OpenStreetMap is a free, online, volunteered geographic information service and database comprised by street-level features and global coverage. Because of its crowd-sourced nature, the dataset provides more up-to-date and high coverage of the road system than the official datasets provided by the government. 
        
        We chose to focus only on street networks for this study, as they provide the main means of connectivity within the country. The majority of the interurban public transport system consists of bus networks that rely on the underling street network. This public transport system is managed by 214 private companies, includes 3,172 routes. However, a large proportion of Ecuador's settlements lack road-based public transport connectivity. Out of 1,013 considered centres, only 423 (42\%) have regular passenger transport, and many rely on informal transport services or private vehicle usage \cite{benabent2017transporte}. We process the street networks from OpenStreetMap using OSMnx \cite{boeing2017osmnx} to create a weighted primal graph of the road network for the country where nodes represent street intersections and edges street centrelines.
    
        Percolation is applied to this road network following the methodology proposed by Arcaute et al. \cite{arcaute2016cities} as specified in the methods section of this chapter. The percolation on the network corresponds to inducing sub-graphs via a thresholding procedure parametrized by distance. The procedure is defined in terms of a distance parameter that determines clusters of nodes in which node neighbours are preserved only if they are reachable within a certain distance. The distance thresholds considered in the percolation procedure are between $100$ and $12,000$ metres at $10$ metre intervals. 
        
        The results are analysed by looking at the evolution of the largest connected component at the different distance thresholds where different transitions can be observed. These transitions define critical distances for the system and are used to build a hierarchical tree of relevant largest clusters. These will be used to define relevant spatial scales of analysis and to calculate the multiscalar segregation as defined by our framework.
    
        For the road network of Ecuador, we constructed a street network graph composed of 387,797 nodes and 1,027,462 edges from OSM data. The results of the evolution of the largest connected component through percolation can be seen in Figure \ref{fig:percolation}.
    
        The first significant transition of the network is detected at $d_c=200m$ and relates to most major cities in the country. This transition is therefore representative of a system of cities. The giant cluster in the next transition of $d_c=1,000m$ corresponds to Quito's metropolitan region. At the transition of $d_c=4,000m$ the division between the coastal areas, the cities in the Andean highlands, and the Amazon can be observed. At the next threshold of $d_c = 6,000m$ the south Andean region merges with the coastal region and form the largest connected component, and at $d_c = 8,000$ both Andean region and Coastal region merge, and only areas in the Amazon near the border with Peru and Colombia are separated. Figure \ref{fig:hierarchical_tree} shows the hierarchical tree formed by the relevant clusters at each scale.
    
    \subsection*{Hierarchical structure of segregation in Ecuador}

        \begin{figure}[hbt!]
            \centering
            \includegraphics[width=1\linewidth]{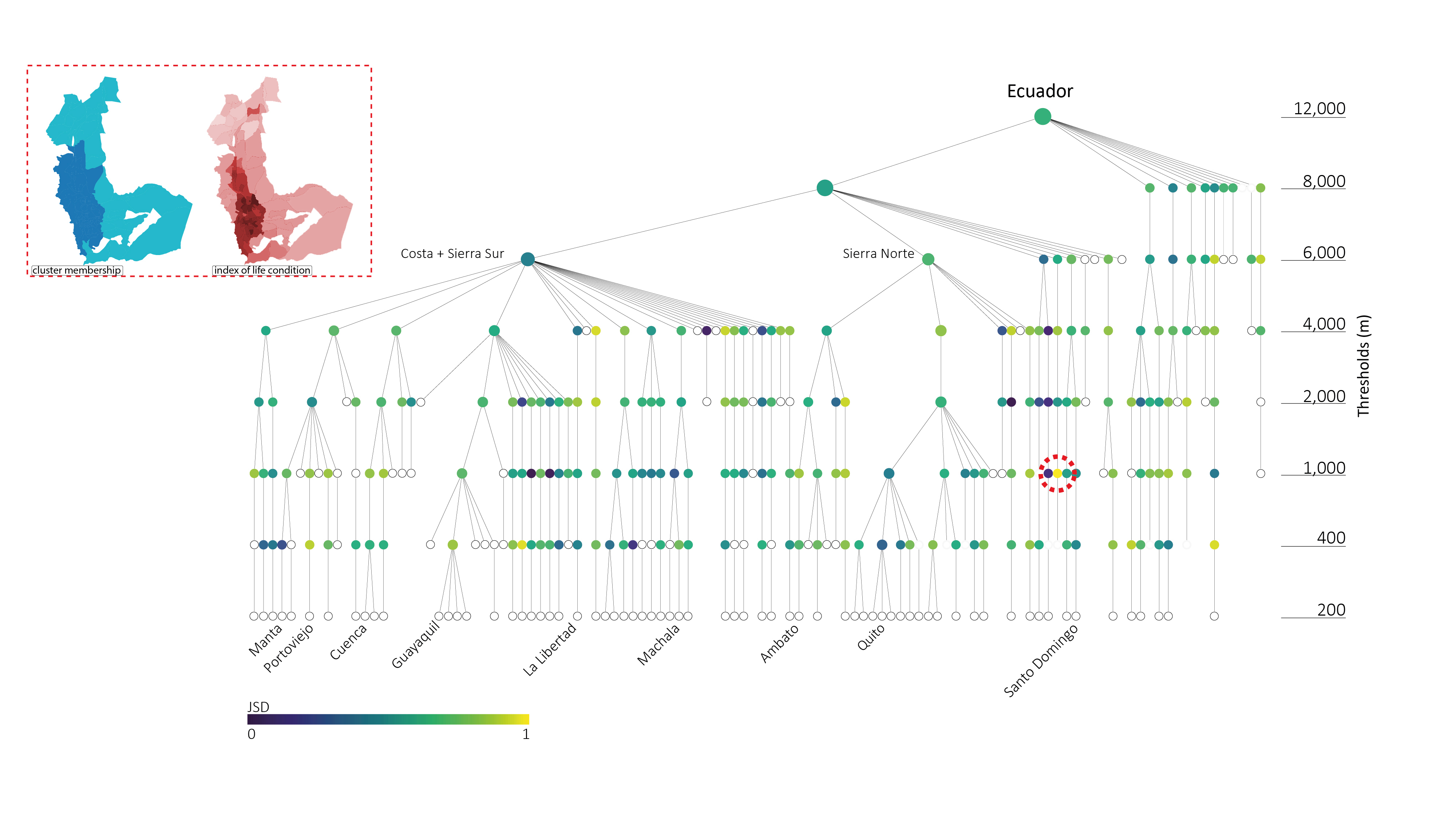}
            \caption[Hierarchical segregation measure through the GJSD index for each parent node in the hierarchical tree of connectivity of Ecuador.]{\label{fig:hierarchical_tree_GJSD} Hierarchical segregation measure through the GJSD index for each parent node in the hierarchical tree of connectivity. The region with the highest segregation value is highlighted and a map of both cluster membership and index of life conditions is shown in the top inset. This region is where the city of Guaranda, in central Ecuador, lies along with its wider region.}
        \end{figure}

        \begin{figure}[hbt!]
            \centering
            \includegraphics[width=0.9\linewidth]{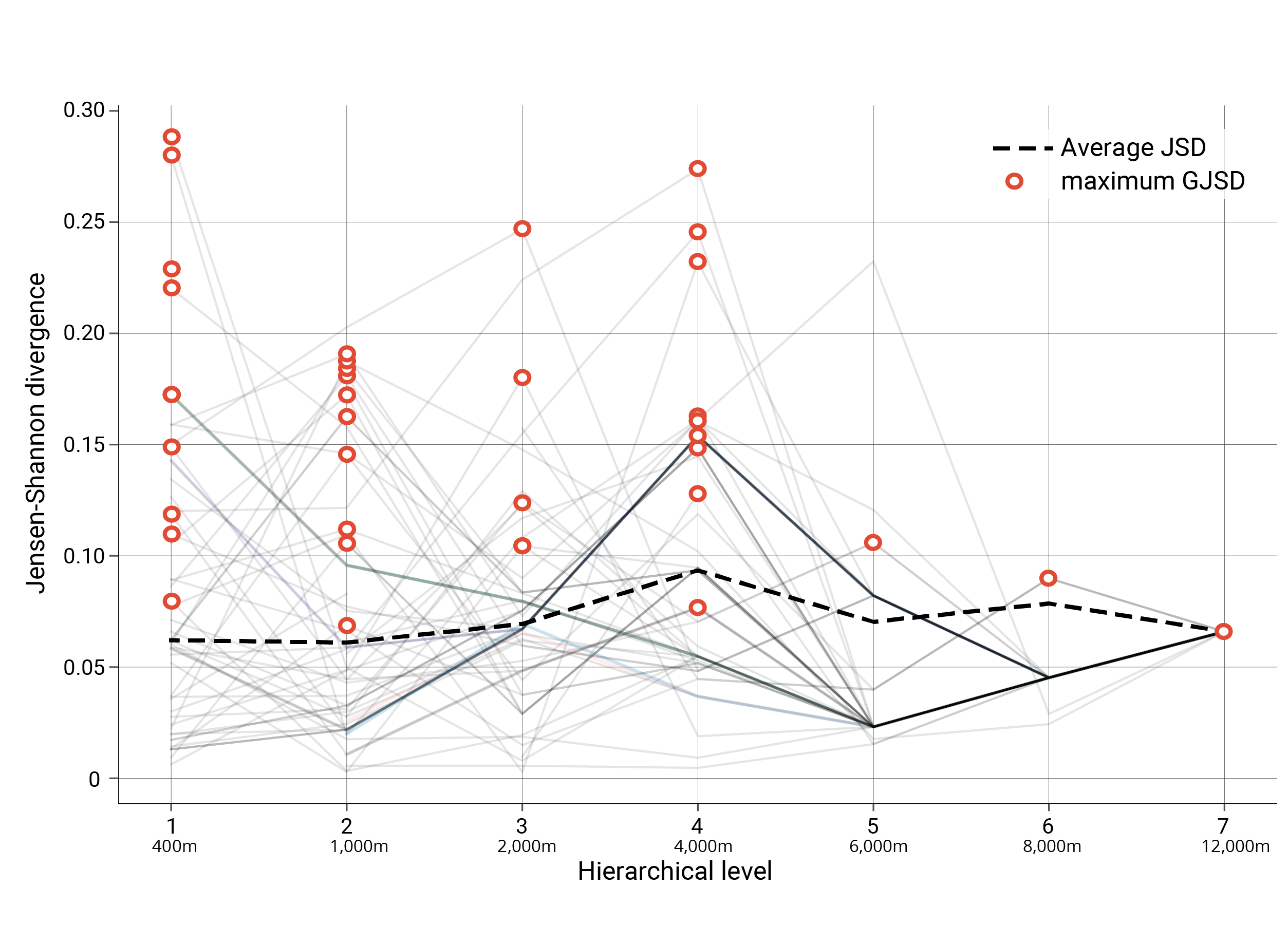}
            \caption[Change in segregation for each city through the hierarchical tree.]{\label{fig:GJSD_local_maximums} Change in segregation for each city as we move up the hierarchical tree. Each line in the graph represents a city in Ecuador. The red dots highlights the relevant scale at which segregation takes place, given by: $\Upsilon_{i}=\argmax\limits_{\delta} \left\lbrace \text{JSD}_{\pi_{\text{norm}}}(i,\delta)\right\rbrace$  }
        \end{figure}
    
        \begin{figure}[hbt!]
            \centering
            \includegraphics[width=0.9\linewidth]{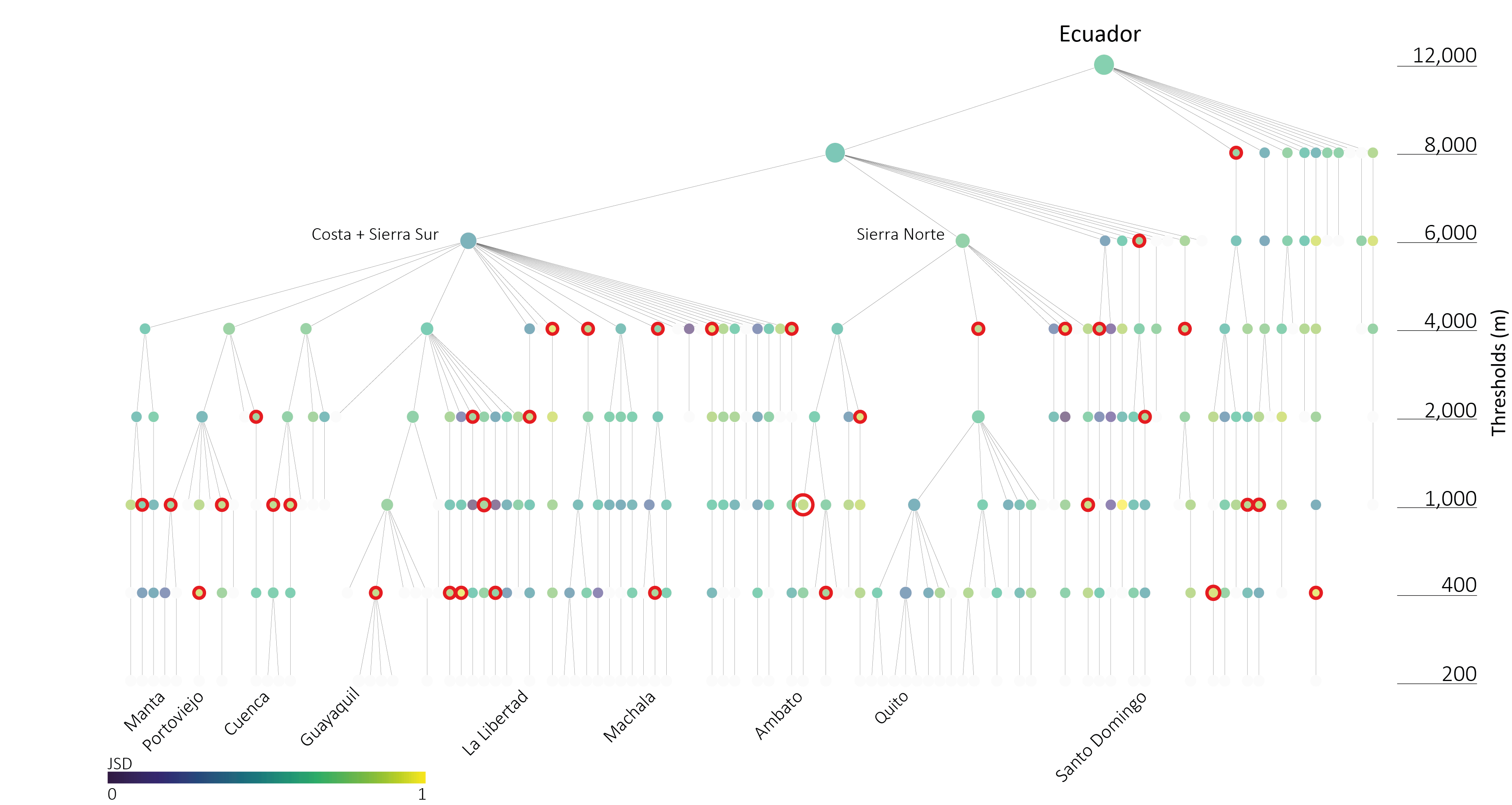}
            \caption[Relevant scales of segregation in Ecuador]{\label{fig:hierarchical_tree_GJSDmax} Relevant scales at which segregation takes places. In the case of the City of Guayaquil, segregation is most pronounced within the city and its periurban area. Quito, on the other hand - presents higher levels of segregation at a much bigger spatial scale, forming a mega-region with several urban centres.}
        \end{figure} 
    
        Using both the hierarchical tree derived through the percolation process on the road network and the index of life conditions derived from the census data - we quantify changes in segregation as we move up through the different spatial scales defined by the hierarchical tree. Figure \ref{fig:hierarchical_tree_GJSD} allows us to visualise each parent node in our hierarchical tree coloured by its normalised GJSD value. 
        
        In the context of Guanujo and Guaranda, two districts within the Bolivar province of Ecuador, our analysis method reveals insightful patterns of segregation, particularly between urban and rural areas. This segregation is highlighted by the highest normalised Generalised Jensen-Shannon Divergence (GJSD) value observed at the 1km threshold within our hierarchical tree model. This finding is especially relevant given Ecuador's diverse topography and the economic significance of agriculture in regions like Guaranda. The method's sensitivity to local scale segregation is crucial in areas where urban-rural divides are pronounced, a common characteristic in agricultural economies.
    
        Moreover, the ability of our method to transcend administrative boundaries, such as those of districts, is a notable strength. It captures the real-world dynamics of segregation that extend beyond political divisions, a crucial aspect in understanding the spatial organization of different socio-economic groups, such as landowners and agricultural workers. In this specific instance, the method uncovers segregation processes between the urbanized city of Guaranda and its surrounding rural areas, despite their geographical proximity.  The economy of the city in this region is mainly based on agriculture, producing items like fruits, wheat, and sugarcane. The process of segregation is mainly driven by the spatial organisation of landowners and agricultural workers.
    
        We can identify the relevant scales at which different processes of segregation are taking place by traversing up the tree for each city and identifying at which scale the population's difference in terms of quality of life is highest. In Figure \ref{fig:GJSD_local_maximums} we see that there is no one particular spatial scale at which segregation takes place, but rather that it is a multiscalar phenomena shaped by overlapping social, economic, and infrastructural processes that operate at and between different geographic levels. For example, for some areas segregation manifests spatially within city regions - such is the case for Guayaquil. In other regions it manifests at a wider regional scale between different Cantons (administrative divisions formed by districts) or Provinces (administrative divisions formed by cantons) - for example in Quito and its wider metropolitan area, see Figure \ref{fig:hierarchical_tree_GJSDmax}). 
    
        Averaging the normalised GJSD segregation values (represented by the dashed line in Figure \ref{fig:GJSD_local_maximums}) at each spatial threshold, we find that for the entire urban system, segregation is maximal at the 4km threshold. This threshold roughly corresponds to provinces in the region - although in some cases the areas the clusters are either smaller then the provinces or cut through multiple ones. This is due to higher differences in the distribution of the underlying population in each of the sub-regions that are identified at the lower threshold of 2km - roughly corresponding to the cantons in the region.

     These results together give initial insights into both the connectivity of different regions and of overall segregation patterns across the country. In the case of the connectivity of the country, we analysed the hierarchical structure of the street network by applying percolation theory. It is surprising to note that most cities remain as independent components of the network until a threshold of around $d=4000m$, where major geographical regions can be observed; such as coastal areas, highland regions, and the Amazon - it is also at this scale where segregation is most pronounced. Even though at these higher scales we would expect to see these divisions because of the topographic characteristics of country, at lower thresholds further research is needed to understand why cities remain relatively disconnected from each other, and how this affects spatial inequality patterns. 
    
\section*{Discussion}

    This study presents a novel analytical framework to analyse socio-spatial segregation at various spatial scales. By incorporating factors of regional connectivity and spatial population distribution, this approach unlocks a solid understanding of segregation pattern shifts across scales. The application of this framework using Ecuador as a case study demonstrates its usefulness in identifying segregated and disconnected regions at differing scales, thus offering consequential insights for future planning and policy-making. Despite the challenges present in understanding the multiscalar nature of socio-spatial segregation, the framework presented here brings about an enhanced methodological approach to facilitate further research in this field.

    The framework addresses the multifaceted problem of socio-spatial segregation. By examining spatial scales using this nested analytical framework, the study can leverage the high granularity of socio-economic data across regions and sub-regions. This method critically considers, not just the spatial distribution of the population, but also the connectivity of these populations within and between regions, contributing a novel perspective to segregation study.

    Our results demonstrate that socio-spatial segregation is neither a phenomenon of purely neighbourhood-scale interaction nor one that is adequately captured by city-wide metrics. Instead, it emerges from the continuous superposition of processes acting at and between distinct spatial scales. By coupling an information-theoretic measure of population heterogeneity to a percolation-derived hierarchy of the street network, we revealed a set of latent "scales of segregation".  For Ecuador, the strongest discontinuity appears at $\delta\approx4\text{ km}$, corresponding roughly to the transition between cantonal and provincial governance.  At this scale, the Andean, Coastal and Amazonian macro-regions disconnect from one another, and the divergence in life-condition is maximised.  Below this threshold, segregation manifests primarily within metropolitan boundaries (e.g. Guayaquil and its peri-urban belt), whereas above it the dominant driver is inter-regional isolation (e.g. Quito’s emerging mega-region and the northern Amazon).  These insights would remain invisible to classic single-scale indices and other multiscale measures that don't account for the underlying connectivity of the regions.

    The identification of scale-dependent segregation has direct consequences for policy design. First, it cautions against interventions that only focus on one spatial scale: a housing subsidy effective at the city block may prove ineffective, or even counter-productive, when the dominant segregation patterns lies between municipalities. Second, the framework highlights the role of connectivity in reinforcing or mitigating inequality. Provinces that remain disconnected in the percolation tree until large \(\delta\) thresholds (most notably parts of the Amazon) also exhibit the highest normalised GJSD values, signalling that poor infrastructure is interlinked with socio-economic segregation. Upgrading inter-provincial links therefore offers two advantages: physical integration and a potential reduction in social-spatial segregation across social groups.
    
    In addition to its utility in socio-economic segregation, this method's adaptability extends to a variety of challenges that involve discerning significant cutoff points within hierarchical trees. This attribute is particularly valuable in the analysis of complex systems characterised by hierarchical structures. The method is uniquely equipped to consider not just the structural layout of these trees but also the specific characteristics inherent to each node. Depending on the study's context, these characteristics could range from demographic details and economic indicators to ecological data.

\bibliography{references}

\section*{Acknowledgements}

This work was supported by The Alan Turing Institute’s Turing studentship scheme.

\section*{Author contributions statement}

M.N. developed the theoretical framework, designed the experiments and conducted the analysis. V.M., and E.A. contributed to the theoretical framework. M.N. wrote the paper, and all authors contributed to the structure and fine tuning of the manuscript.


\section*{Competing interests} 
The authors declare no competing interests.

\clearpage

\section*{Supplementary materials}
\label{Supplementary material}

    \subsection{Index of life conditions: ICV}

    Table \ref{table:sup_icv_table} shows the variables used in the calculation of the index of life conditions, along with their weighting.

    \begin{table}[!ht]
    \small
        \centering
            \begin{tabular}{
                 l 
                 l 
                 l 
                 l 
                }

                \textbf{Dimension} & \textbf{Subdimension} & \textbf{Variable}  & \textbf{Range}\\

                Housing  & Quality of housing & Quality of floors &  0-2 \\
                                             && Quality of exterior walls & 0-3 \\
                                             && Quality of roof & 0-2 \\
                        & Quantity (spaces) & Exclusive kitchen &  0-1 \\
                                             && Exclusive bathroom & 0-1 \\
                                             && Number of bedrooms per person & 0-1 \\
                                             && Extra rooms & 0-1 \\
                Basic services  & Water and sanitation & Availability of water &  0-3 \\
                                             && Availability of drainage & 0-2 \\
                                             && Solid waste collection Q& 0-1 \\
                        & Electricity & Availability of electricity &  0-1 \\
                                             && Availability of fuel or energy for cooking & 0-1 \\
                        & Telecommunications & Availability of telephone services &  0-1 \\
                                             && Availability of internet connection & 0-1 \\
                                             && Availability of cable TV & 1-2 \\
                Education  &                 & Years of schooling & 0-1 \\
                Healthcare access &          & Access to health insurance &  0.5-1 \\
                                                
                \end{tabular}
        \caption{Variables used from census to calculate index of life conditions along with the score range. The main dimensions considered for the index are: housing in both terms of quality and quantity; basic services such as water, sanitation, electricity, and telecommunications; education; and healthcare. Income is not considered within the indicator as there is no readily available data. All indicators are defined at the household level, assuming equal sharing and externalities, and are aggregated to the urban block level by taking the average of all the households that belong to a specific block, where a block is defined by areas enclosed by streets, providing a disaggregated view of spatial distribution patterns. After aggregation the data is divided into quartiles.}
        \label{table:sup_icv_table}
    \end{table}

    \begin{figure}
        \centering
        \includegraphics[width=0.9\linewidth]{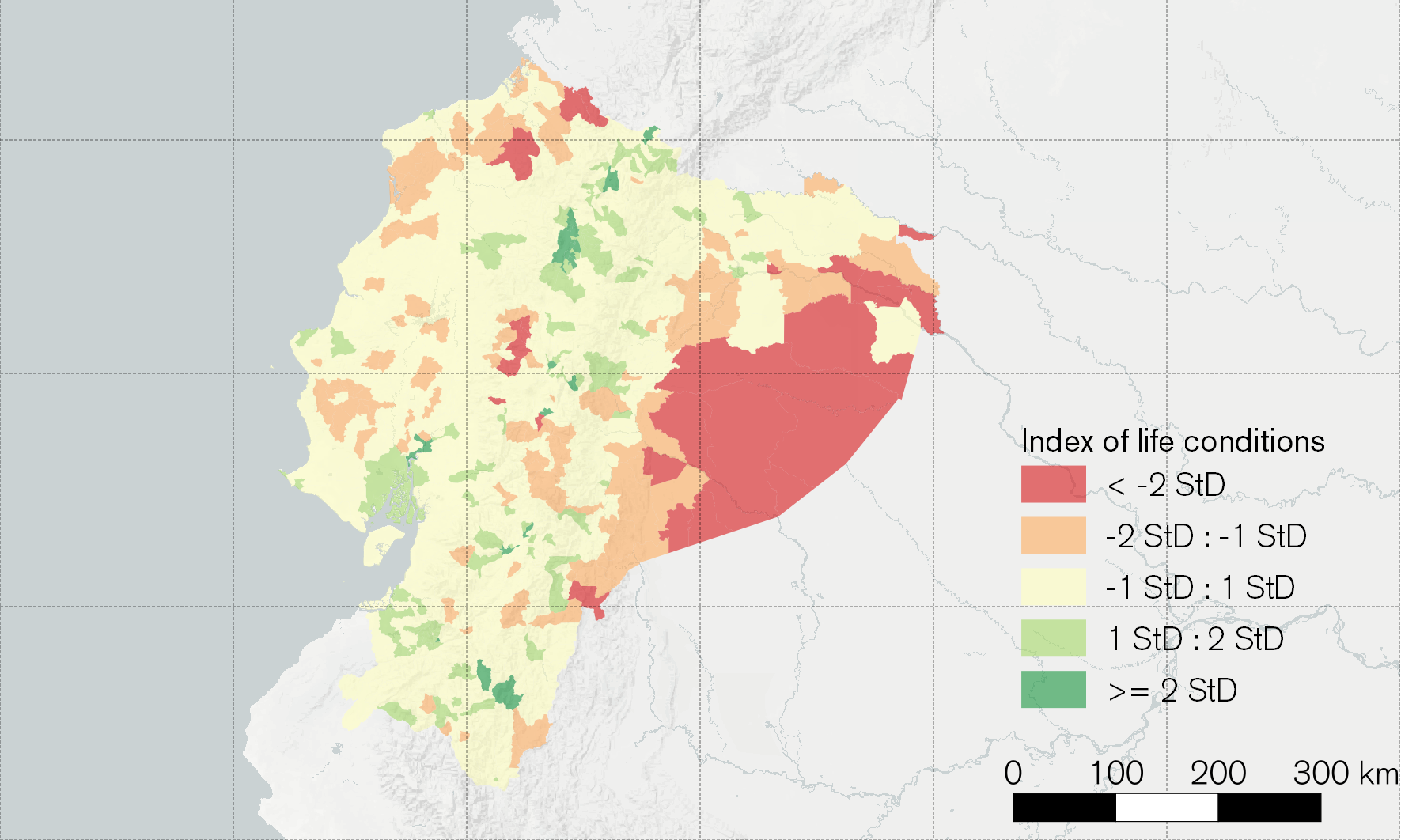}
        \caption[Spatial distribution of quality of life index for Ecuador, calculated from census data.]{\label{fig:icv_map_ec} Spatial distribution of the deviations of the national average of the quality of life index ICV for Ecuador (calculated from census data). We observe that the population with high standard of life conditions are located around major urban centres like Quito, Guayaquil, and Cuenca. The population with the lowest life conditions are situated mainly in rural areas and in Amazonic region of the country.}
    \end{figure}

\begin{figure}
        \centering
        \includegraphics[width=0.9\linewidth]{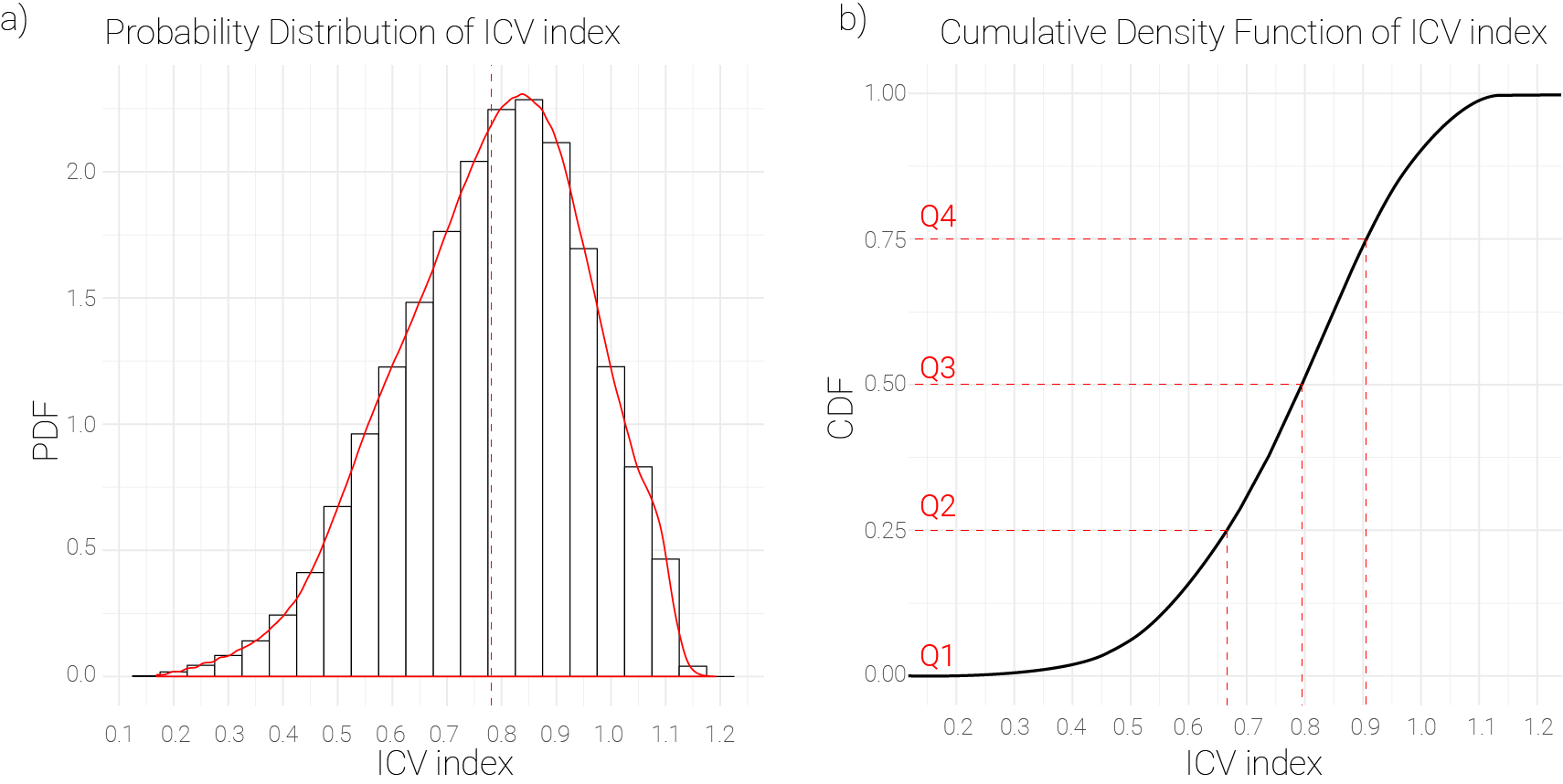}
        \caption[Statistical distribution of index of life conditions for Ecuador, calculated from census data]{\label{fig:icv_stats_ec} Distribution of index of life conditions for Ecuador, calculated from census data 2010. Values of 1 and above represent no deprivation along any of the dimensions included in the index.}
    \end{figure}

\begin{figure}
        \centering
        \includegraphics[width=0.9\linewidth]{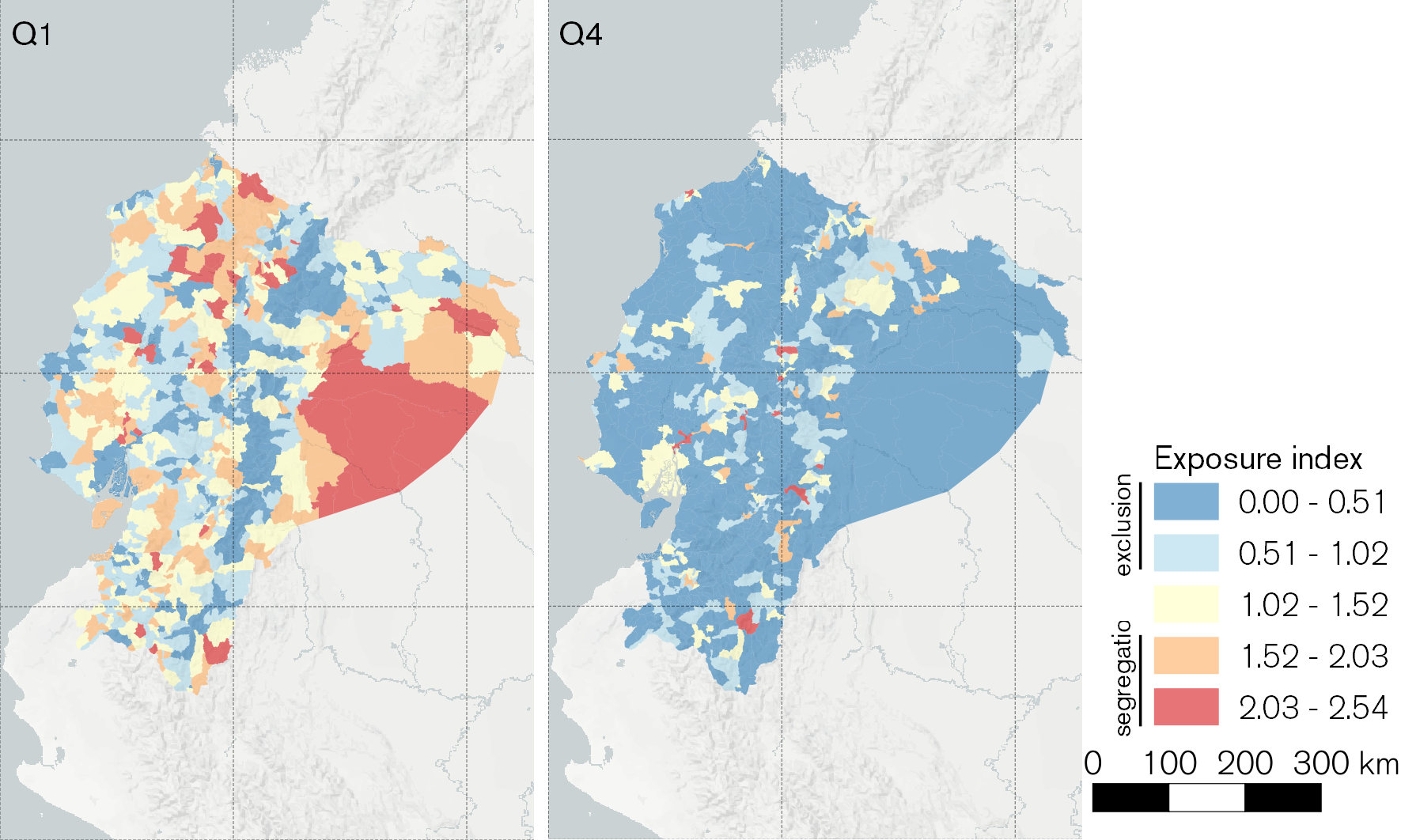}
        \caption[Exposure index for the lowest and highest quartiles of the Index of Life Conditions values in Ecuador.]{\label{fig:exposure_ec} Exposure index for the lowest and highest quartiles of the Index of Life Conditions values. Both processes of exclusion and segregation can be identified for both the highest and lowest quartiles of the population.}
    \end{figure}

    The index of life conditions $ICV$ was calculated for the population at block level using census data from 2010. Figure \ref{fig:icv_stats_ec} shows the distribution of the index values for the whole population. The country has a mean value of 0.78 and a median value of 0.85. Values of 1 and above represent no deprivation along any of the dimensions included in the index. The index of life conditions was aggregated at the \emph{Parroquia} level, which corresponds to administrative boundaries that are smaller than metropolitan regions. The aggregation is done by averaging the ICV values calculated at the household level of all households that belong to each Parroquia. The spatial distribution of the deviations from the national average of the ICV value at the Parroquia level is shown in Figure \ref{fig:icv_map_ec}. Ecuador contains a total of $1499$ Parroquias, where the ones containing major cities like Quito, Guayaquil, and Cuenca, exhibit an index of life conditions with 1 std deviation above the national average, while the areas with the lowest values are concentrated in the Amazon region near the border with Peru. Using these values, the population was divided into quartiles, and the quartiles where aggregated at the scale of Parroquia to analyse its spatial distribution. We look at the lowest and highest quartiles and plot their exposure within each region, as shown in Figure \ref{fig:exposure_ec}. Exposure is measured through the ratio of the relative size of one population group in the area  $b_i$  and the relative size of the total population in the same area $a_i$: $exposure = b_i/a_i$. Low exposure values indicate areas of exclusion, where the population of a specific quartile is underrepresented. High exposure values, conversely, denote areas of segregation, where the population of a quartile is over-represented. The lowest quartile are excluded from major cities, and exhibit patterns of both exclusion and segregation, while the highest quartile mainly exhibit patterns of exclusion, with segregation in very few small areas.

    In this context, exposure values serve as indicators of societal dynamics, specifically in terms of how population groups are distributed across different areas. Exclusion is evident when a particular population group, represented by a quartile, is less prevalent in an area compared to their representation in the overall population. This underrepresentation, marked by low exposure values, suggests barriers that prevent the group from being present or living in those areas. Conversely, segregation is observed when a population group is overly concentrated in a specific area, as indicated by high exposure values. This overrepresentation implies that the group is more isolated or clustered, likely due to various social, economic, or political factors. 
            
    We have shown segregation patterns for the entire country using an index of life conditions as a proxy for different socio-economic groups. Although information exists for spatial distribution of the population in the country, existing research has only looked at distributions using hard thresholds dividing the population into groups that are above or below the poverty line. By using a composite index we have shown a more detailed picture of how the population is distributed according to their socio-economic attributes. As expected, areas with higher life conditions are located in urbanised areas and bigger cities, such as Quito, Guayaquil, and Cuenca; as people in these areas have easier access to water and sanitation infrastructure, as well as educational and health facilities. 
    
    Interestingly, when looking at spatial segregation patterns for the lowest and highest quartiles of the population distributions, we see two different patterns of segregation emerging. In the case of the group in the highest quartile, these are concentrated in few small geographical areas, and mostly underrepresented in the rest of the Country. In contrast, the group in the lowest quartile is more evenly distributed across the region, with areas showing patterns of exclusion along the Andean region, and some coastal areas. 
    
    This spatial segregation patterns hint at the existence of possible self-segregation dynamics in the case of the highest quartiles, and exclusion in the lowest. One of the main forms of self-segregation in Ecuador are gated communities; residential areas with restricted access and often enhanced security. In Ecuador, particularly in cities coastal cities, but also through the entire country, these gated communities have become increasingly prevalent, symbolising a physical and social divide between different socioeconomic groups \cite{perez2023geografia}.

    \subsection*{Formal Derivation of the Maximum Value of GJSD}
        Here, we provide a more formal derivation to show why the maximum value of the Generalised Jensen-Shannon Divergence (GJSD) for a set of weighted distributions is equal to the entropy of the weight distribution $\pi$. \\
        
        Consider a set of distributions $P_1, P_2, \ldots, P_n$ defined over a common discrete support, and let $\pi = (\pi_1, \pi_2, \ldots, \pi_n)$ be a probability weight vector with $\pi_i \geq 0$ and $\sum_{i=1}^n \pi_i = 1$. Define the mixture distribution:

        \begin{equation}
            M = \sum_{i=1}^n \pi_i P_i.
        \end{equation}
        
        The Generalised Jensen-Shannon Divergence is given by:
        \begin{equation}
        \label{eq:GJSD_sup}
        \text{JSD}_{\pi}(P_1, P_2, \ldots, P_n) = H(M) - \sum_{i=1}^n \pi_i H(P_i),
        \end{equation} 
        
        where $H(X) = -\sum_{x} p(x)\log p(x)$ is the Shannon entropy. \\
        
        We seek to identify the maximum value of $\text{JSD}_{\pi}$. Maximising $\text{JSD}_{\pi}$ involves finding the set of distributions $\{P_i\}$ that yield the greatest possible difference between $H(M)$ and $\sum_{i} \pi_i H(P_i)$. \\
        
        To maximise $\text{JSD}_{\pi}$, we consider the scenario where each $P_i$ is as distinct as possible from the others. Specifically, let each $P_i$ be a probability distribution concentrated entirely on a single, unique element $ x_i $. Formally:

        \begin{equation}
            P_i(x) = 
            \begin{cases}
            1 & \text{if } x = x_i \\[6pt]
            0 & \text{otherwise.}
            \end{cases}
        \end{equation}
        
        Furthermore, assume the supports are mutually exclusive, i.e., $x_i \neq x_j$ for all $i \neq j$. This ensures no overlap between any two distributions.\\
        
        \noindent \textbf{Entropy of Each $P_i$:}  
        Since $P_i$ is concentrated on a single outcome, it has zero entropy:
        \begin{equation}
            \label{eq:Pi_zero}
            H(P_i) = 0.
        \end{equation} \\
        
        \noindent \textbf{Entropy of the Mixture $M$:}  
        Under these conditions, the mixture $M$ combines these distinct elements, each associated uniquely with one distribution:
        \begin{equation}
            \label{eq:shannon_mixure}
        M(x) = \pi_i \quad \text{if } x = x_i, \quad \text{and } M(x) = 0 \text{ otherwise.}
        \end{equation}
        
        Thus, $M$ is itself a discrete distribution over the set $\{x_1, x_2, \ldots, x_n\}$ with probabilities given by $\pi_i$. Its entropy is:
        \begin{equation}
            H(M) = -\sum_{i=1}^n \pi_i \log \pi_i.
        \end{equation}\\
        
        \noindent \textbf{Maximum GJSD:} Substituting eq. \ref{eq:Pi_zero} and eq. \ref{eq:shannon_mixure} back into the GJSD equation \ref{eq:GJSD_sup} we get:

        \begin{equation}
            \text{JSD}_{\text{max}}(P_1, \ldots, P_n) = H(M) = -\sum_{i=1}^n \pi_i \log \pi_i.
        \end{equation}
        
        This value depends solely on the distribution of the weights $\pi$. Since this construction corresponds to a scenario in which all distributions are perfectly distinct, it represents the maximal dissimilarity scenario. Any other configuration of distributions, where some overlap occurs, will result in strictly lower $\text{JSD}_{\pi}$ values, as partial overlaps decrease the difference between $H(M)$ and $\sum \pi_i H(P_i)$.

\end{document}